\documentclass[english]{svmult}
\usepackage[T1]{fontenc}
\usepackage[latin9]{inputenc}
\usepackage{geometry}
\usepackage{float}
\usepackage{amsmath}
\usepackage{amssymb}
\usepackage{graphicx}

\usepackage{mathptmx,helvet,courier,graphicx,makeidx,multicol,footmisc}

\makeatletter

\providecommand{\tabularnewline}{\\}

\usepackage[T1]{fontenc} 

\PassOptionsToPackage{normalem}{ulem} 
\usepackage{babel}
\usepackage{cite}
\usepackage{color}
\usepackage{dsfont}
\usepackage{empheq}
\usepackage{framed}
\usepackage{mathrsfs}
\usepackage{needspace}
\usepackage{setspace}
\usepackage{tabularx}
\usepackage{tikz}
\usepackage{ulem}
\usepackage{url}

\newcommand{\mm}{\mu_{\mathrm{macro}}}
\newcommand{\lm}{\lambda_{\mathrm{macro}}}
\newcommand{\mh}{\mu_{\mathrm{micro}}}
\newcommand{\lh}{\lambda_{\mathrm{micro}}}
\newcommand{\me}{\mu_{e}}
\newcommand{\mc}{\mu_{c}}
\newcommand{\lle}{\lambda_{e}}

\newcommand{\mLc}{\me L_{c}^{2}}
\newcommand{\mLd}{\me L_{d}^{2}}

\newcommand{\R}{\mathbb{R}}
\newcommand{\nablau}{\,\nabla u\,}
\newcommand{\p}{{P}}
\newcommand{\nablap}{\nabla \p}
\newcommand{\Curl}{\,\mathrm{Curl}}
\newcommand{\dev}{\, \mathrm{dev}}
\newcommand{\Div}{\mathrm{Div}}
\newcommand{\tr}{\, \mathrm{tr}}
\newcommand{\sym}{\, \mathrm{sym}\,}

\renewcommand{\skew}{\, \mathrm{skew}\,}

\renewcommand{\skew}{\, \mathrm{skew}}

\newcommand{\id}{\,\mathds{1}}

\definecolor{Green}{rgb}{0,0.52,0}

\makeatother

\usepackage{babel}
\begin{document}

\title*{A review on wave propagation modeling in band-gap
metamaterials via enriched continuum models}

\author{Angela Madeo and Patrizio Neff and  Gabriele Barbagallo and Marco Valerio d'Agostino and	Ionel-Dumitrel Ghiba}

\institute{Angela Madeo \at LGCIE,INSA-Lyon, Universitè de Lyon, 20 avenue Albert Einstein, 69621, Villeurbanne cedex and IUF, Institut universitaire de France, 1 rue Descartes, 75231 Paris Cedex 05, France, 	\email{angela.madeo@insa-lyon.fr} 
	\and Patrizio Neff \at Head of Chair for Nonlinear Analysis and Modelling, Fakultät für Mathematik, Universität Duisburg-Essen, Mathematik-Carrée, Thea-Leymann-Straße 9, 45127 Essen, Germany, \email{patrizio.neff@uni-due.de} 
\and Gabriele Barbagallo \at LaMCoS-CNRS \textbackslash{}\& LGCIE, INSA-Lyon, Universitité de Lyon, 20 avenue Albert Einstein, 69621, Villeurbanne cedex, France, \email{gabriele.barbagallo@insa-lyon.fr} 
\and Marco Valerio d'Agostino \at LGCIE, INSA-Lyon, Université de Lyon, 20 avenue Albert Einstein, 69621, Villeurbanne cedex, France,  \email{marco-valerio.dagostino@insa-lyon.fr} \and Ionel-Dumitrel Ghiba,  Lehrstuhl für Nichtlineare Analysis und Modellierung, Fakultät für Mathematik, Universität Duisburg-Essen, Thea-Leymann Str.  9, 45127 Essen, Germany; Alexandru Ioan Cuza University of Ia{\c{s}}i, Department of Mathematics, Blvd.  Carol I, no.  11, 700506 Ia{\c{s}}i, Romania; and Octav Mayer Institute of Mathematics of the Romanian Academy, Ia{\c{s}}i Branch, 700505 Ia{\c{s}}i, \email{dumitrel.ghiba@uni-due.de,  dumitrel.ghiba@uaic.ro}}

\maketitle

\abstract{In the present contribution we show that the relaxed micromorphic model is the only non-local continuum model which is able to account for the description of band-gaps in metamaterials for which the kinetic energy accounts separately for micro and macro-motions without considering a micro-macro coupling. Moreover, we show that when adding a gradient inertia term which indeed allows for the description of the coupling of the vibrations of the microstructure to the macroscopic motion of the unit cell, other enriched continuum models of the micromorphic type may allow the description of the onset of band-gaps. Nevertheless, the relaxed micromorphic model proves to be yet the most effective enriched continuum model which is able to describe multiple band-gaps in non-local metamaterials.}


\section{Introduction}

In the last years, a lot of interest has been raised from a class of microscopically heterogeneous materials which show exotic behaviors such as that of ``stopping'' the propagation of elastic waves. In some cases, the waves lose some of the energy due to micro-diffusion phenomena (Bragg scattering) or even local resonance of the microstructure (Mie resonance). These effects can be exploited to design innovative materials whose dynamical behavior differs completely from the classical materials usually employed in engineering sciences.

As a matter of fact, classical Cauchy continuum theories, are not always well adapted to cover the wealth of experimental evidences on the dynamical behavior of real materials. As a first point, in fact, real materials commonly show dispersive behaviors, which means that the speed of propagation of the traveling wave changes when considering smaller wavelengths. Such phenomenon is not astonishing if one thinks that the structure of matter changes when observing it at smaller scales. It suffices to go down to the scale of the crystals or molecules to be aware of the heterogeneity of matter. It is for this reason that waves with wavelengths which are small enough to ``sense'' the presence of the microstruture will propagate at a different speed than other waves with higher wavelengths. Cauchy continuum theories are not able in any case to account for dispersive phenomena and are a good approximation of reality only for those materials which do not exhibit their heterogeneity at the scale of interest. As far as one wants to model dispersive behaviors, Cauchy continuum theories are no longer adapted and more refined models need to be introduced. One possibility is to introduce second or higher order theories so allowing the description of dispersion for the acoustic modes (see e.g. \cite{dellisola2012linear,placidi2014reflection}). Nevertheless, if second gradient theories may, on the one hand, be of use for the description of some dispersive behaviors, on the other hand they are often insufficient to describe more complex behaviors of metamaterials in which the microstructure can have its own vibrational modes independently of the motion of the unit cell. In order to describe the complex dynamical behavior of such metamaterials in a continuum framework, the introduction of continuum models with enriched kinematics (micromorphic models) is a mandatory requirement \cite{mindlin1963microstructure,eringen1999microcontinuum,madeo2014band,madeo2015wave,madeo2016complete}. Continuum models of the micromorphic type, in fact, allow for the description of microstructure-related vibrational modes thanks to the introduction of extra degrees of freedom with respect to the displacement field alone. 

The insufficiency of Cauchy continuum theories becomes even more evident when considering more complex metamaterials which are able to inhibit wave propagation, i.e. so called band-gap metamaterials. To catch the complex dynamical behavior of such materials, even some of the available micromorphic models are not sufficiently adapted. Indeed, it has been shown in previous contributions that the relaxed micromorphic model is the only continuum model of the micromorphic type which is able to account for the description of band-gaps when considering a kinetic energy in which the macroscopic and microscopic motions are completely uncoupled \cite{madeo2014band,madeo2015wave,madeo2016complete,madeo2016reflection}. In this contribution we will show, following what done in \cite{madeo2016role}, that the addition of kinetic energy terms which couple the motions of the microstructure to the macro-motions of the unit cells may have a deep impact on the ability of describing band-gaps behaviors. 

\subsection{Notations}

In this contribution, we denote by $\R^{3\times3}$ the set of real
$3\times3$ second order tensors, written with capital letters. We
denote respectively by $\cdot\:$, $:$ and $\left.\langle\cdot,\cdot\right.\rangle$
a simple and double contraction and the scalar product between two
tensors of any suitable order\footnote{For example, $(A\cdot v)_{i}=A_{ij}v_{j}$, $(A\cdot B)_{ik}=A_{ij}B_{jk}$,
$A:B=A_{ij}B_{ji}$, $(C\cdot B)_{ijk}=C_{ijp}B_{pk}$, $(C:B)_{i}=C_{ijp}B_{pj}$,
$\left.\langle v,w\right.\rangle=v\cdot w=v_{i}w_{i}$, $\left.\langle A,B\right.\rangle=A_{ij}B_{ij}$
etc.}. Everywhere we adopt the Einstein convention of sum over repeated
indices if not differently specified. The standard Euclidean scalar
product on $\R^{3\times3}$ is given by $\langle{X},{Y}\rangle_{\R^{3\times3}}=\tr({X\cdot Y^{T}})$,
and thus the Frobenius tensor norm is $\|{X}\|^{2}=\langle{X},{X}\rangle_{\R^{3\times3}}$.
In the following we omit the index $\R^{3},\R^{3\times3}$. The identity
tensor on $\R^{3\times3}$ will be denoted by $\id$, so that $\tr({X})=\langle{X},{\id}\rangle$. 

\medskip{}

We consider a body which occupies a bounded open set $B_{L}$ of the
three-dimensional Euclidian space $\R^{3}$ and assume that its boundary
$\partial B_{L}$ is a smooth surface of class $C^{2}$. An elastic
material fills the domain $B_{L}\subset\R^{3}$ and we refer the motion
of the body to rectangular axes $Ox_{i}$. 

For vector fields $v$ with components in ${\rm H}^{1}(B_{L})$, i.e.
$v=\left(v_{1},v_{2},v_{3}\right)^{T}\,,v_{i}\in{\rm H}^{1}(B_{L}),$
we define \break $\nabla\,v=\left((\nabla\,v_{1})^{T},(\nabla\,v_{2})^{T},(\nabla\,v_{3})^{T}\right)^{T}$,
while for tensor fields $P$ with rows in ${\rm H}({\rm curl}\,;B_{L})$,
resp. ${\rm H}({\rm div}\,;B_{L})$, i.e. \break $P=\left(P_{1}^{T},P_{2}^{T},P_{3}^{T}\right)$,
$P_{i}\in{\rm H}({\rm curl}\,;B_{L})$ resp. $P_{i}\in{\rm H}({\rm div}\,;B_{L})$
we define ${\rm Curl}\,P=\left(({\rm curl}\,P_{1})^{T},({\rm curl}\,P_{2})^{T},({\rm curl}\,P_{3})^{T}\right)^{T},$
${\rm Div}\,P=\left({\rm div}\,P_{1},{\rm div}\,P_{2},{\rm div}\,P_{3}\right)^{T}.$ 

A subscript $_{,t}$ will indicate derivation with respect to time and, analogously a subscript $_{,tt}$ stands for the second derivative of the considered quantity with respect to time.

\vspace{0.2cm}

As for the kinematics of the considered micromorphic continua, we
introduce the functions
\[
\chi(X,t):B_{L}\rightarrow\mathbb{R}^{3},\qquad P(X,t):B_{L}\rightarrow\mathbb{R}^{3\times3},
\]
which are known as \textit{placement} vector field and \textit{micro-distortion}
tensor, respectively. The physical meaning of the placement field
is that of locating, at any instant $t$, the current position of
the material particle $X\in B_{L}$, while the micro-distortion field
describes deformations of the microstructure embedded in the material
particle $X$. As it is usual in
continuum mechanics, the displacement field can also be introduced
as the function $u(X,t):B_{L}\rightarrow\mathbb{R}^{3}$ defined as
\[
u(X,t)=\chi(X,t)-X.
\]

\medskip{}

In the remainder of the paper, the following acronyms will be used
to refer to the branches of the dispersion curves: 
\begin{itemize}
\item TRO: transverse rotational optic, 
\item TSO: transverse shear optic, 
\item TCVO: transverse constant-volume optic, 
\item LA: longitudinal acoustic, 
\item LO$_{1}$-LO$_{2}$: $1^{st}$ and $2^{nd}$ longitudinal optic, 
\item TA: transverse acoustic, 
\item TO$_{1}$-TO$_{2}$: $1^{st}$ and $2^{nd}$ transverse optic. 
\end{itemize}
\medskip{}

If not differently specified, the results presented in this paper
are obtained for values of the elastic coefficients chosen as in Table
\ref{ParametersValues} (see Equations \eqref{eq:Kinetic}, \eqref{eq:Ener-General},
\eqref{eq:Ener-DivCurl}, \eqref{eq:Ener-Div}, \eqref{eq:Ener-Mindlin}
and \eqref{eq:Ener-Int} for their definition).

\begin{table}[H]
\begin{centering}
\begin{tabular}{|c|c|c|}
\hline 
Parameter  & Value  & Unit\tabularnewline
\hline 
\hline 
$\me$  & $200$  & $MPa$\tabularnewline
\hline 
$\lle=2\me$  & 400  & $MPa$\tabularnewline
\hline 
$\mc=5\me$  & 1000  & $MPa$\tabularnewline
\hline 
$\mh$  & 100  & $MPa$\tabularnewline
\hline 
$\lh$  & $100$  & $MPa$\tabularnewline
\hline
\end{tabular}\quad{}\quad{}\quad{}\quad{}%
\begin{tabular}{|c|c|c|}
	\hline 
	Parameter  & Value  & Unit\tabularnewline
	\hline  
	\hline 
$L_{c}\ $  & $1$  & $mm$\tabularnewline
\hline 
$\rho$  & $2000$  & $kg/m^{3}$\tabularnewline
\hline 
$\eta$  & $10^{-2}$  & $kg/m$\tabularnewline
\hline 
$\overline{\eta}_{i}$  & $10^{-1}$  & $kg/m$\tabularnewline
\hline 
\end{tabular}\quad{}\quad{}\quad{}\quad{}%
\begin{tabular}{|c|c|c|}
\hline 
Parameter  & Value  & Unit\tabularnewline
\hline 
\hline 
$\lm$  & $82.5$  & $MPa$\tabularnewline
\hline 
$\mm$  & $66.7$  & $MPa$\tabularnewline
\hline 
$E_{\mathrm{macro}}$  & $170$  & $MPa$\tabularnewline
\hline 
$\nu_{\mathrm{macro}}$  & $0.28$  & $-$\tabularnewline
\hline 
\end{tabular}
\par\end{centering}
\caption{\label{ParametersValues}Values of the elastic parameters used in the numerical
simulations (left), characteristic lengths and inertiae (center) and corresponding values of the Lamé parameters
and of the Young modulus and Poisson ratio (right), for the formulas
needed to calculate the homogenized macroscopic parameters starting
from the microscopic ones, see \cite{barbagallo2016transparent}.}
\end{table}

\subsection{The fundamental role of micro-inertia in enriched continuum mechanics}

As far as enriched continuum models are concerned, a central issue
which is also an open scientific question is that of identifying the
role of so-called micro-inertia terms on the dispersive behavior of
such media. As a matter of fact, enriched continuum models usually
provide a richer kinematics, with respect to the classical macroscopic
displacement field alone, which is related to the possibility of describing
the motions of the microstructure inside the unit cell. The adoption
of such enriched kinematics (given by the displacement field $u$
and the micro-distortion tensor $P$, see e.g. \cite{mindlin1964micro,eringen1999microcontinuum,ghiba2014relaxed,madeo2014band,madeo2015wave,madeo2016complete,madeo2016first,madeo2016reflection,neff2014unifying})
as we will see, allows for the introduction of constitutive laws for
the strain energy density that are able to describe the mechanical
behavior of some metamaterials in the static regime. When the dynamical
regime is considered, things become even more delicate since the choice
of micro-inertia terms to be introduced in the kinetic energy density
must be carefully based on 
\begin{itemize}
	\item a compatibility with the chosen kinematics and constitutive laws used
	for the description of the static regime,
	\item the specific inertial characteristics of the metamaterial that one
	wants to describe (e.g. eventual coupling of the motion of the microstructure
	with the macro motions of the unit cell, specific resistance of the
	microstructure to independent motion, etc.).
\end{itemize}
In the present paper, we will suppose that the kinetic energy takes
the following general form (see \cite{madeo2016role}): 
\begin{gather}
	J=\hspace{-0.1cm}\underbrace{\frac{1}{2}\rho\left\Vert u_{,t}\right\Vert ^{2}}_{\text{Cauchy inertia}}+\hspace{-0.1cm}\underbrace{\frac{1}{2}\eta\left\Vert \p_{,t}\right\Vert ^{2}}_{\text{free micro-inertia}}\hspace{-0.3cm}+\hspace{0.1cm}\underbrace{\frac{1}{2}\overline{\eta}_{1}\left\Vert \dev\sym\nablau_{,t}\right\Vert ^{2}+\frac{1}{2}\overline{\eta}_{2}\left\Vert \skew\nablau_{,t}\right\Vert ^{2}+\frac{1}{6}\overline{\eta}_{3}\tr\left(\nablau_{,t}\right)^{2}}_{\text{new gradient micro-inertia}},\label{eq:Kinetic}
\end{gather}
where $\rho$ is the value of the average macroscopic mass density
of the considered metamaterial, $\eta$ is the free micro-inertia
density and the $\overline{\eta}_{i},i=\{1,2,3\}$ are the gradient
micro-inertia densities associated to the different terms of the Cartan-Lie
decomposition of $\nablau$. We will be hence able to explicitly show
which is the explicit role of the gradient micro-inertia on the onset
of band-gaps in continuum models of the micromorphic type. More precisely,
we will highlight which is the specific effect of the introduction
of gradient micro-inertia terms on different enriched continuum models,
namely:
\begin{itemize}
	\item the classical relaxed micromorphic model, 
	\item the relaxed micromorphic model with curvature $\rVert\Div\p\lVert^{2}+\rVert\Curl\p\lVert^{2}$, 
	\item the relaxed micromorphic model with curvature $\rVert\Div\p\lVert^{2}$, 
	\item the standard Mindlin-Eringen model, 
	\item the internal variable model. 
\end{itemize}

\section{The classical relaxed micromorphic model}

Our relaxed micromorphic model endows Mindlin-Eringen's representation
with the second order \textbf{dislocation density tensor} $\alpha=-\Curl\p$
instead of the third order tensor $\nablap$.\footnote{The dislocation tensor is defined as $\alpha_{ij}=-\left(\Curl\p\right)_{ij}=-\p_{ih,k}\epsilon_{jkh}$,
where $\epsilon$ is the Levi-Civita tensor and Einstein notation
of sum over repeated indices is used.} In the isotropic case the energy of the relaxed micromorphic model
reads
\begin{align}
W= & \underbrace{\me\,\lVert\sym\left(\nablau-\p\right)\rVert^{2}+\frac{\lle}{2}\left(\mathrm{tr}\left(\nablau-\p\right)\right)^{2}}_{\mathrm{{\textstyle isotropic\ elastic-energy}}}+\hspace{-0.1cm}\underbrace{\mc\,\lVert\skew\left(\nablau-\p\right)\rVert^{2}}_{\mathrm{{\textstyle rotational\ elastic\ coupling}}}\hspace{-0.1cm}\label{eq:Ener-General}\\
 & \quad+\underbrace{\mh\,\lVert\sym\p\rVert^{2}+\frac{\lh}{2}\,\left(\mathrm{tr}\p\right)^{2}}_{\mathrm{{\textstyle micro-self-energy}}}+\underbrace{\frac{\mLc}{2}\,\lVert\Curl\p\rVert^{2}}_{\mathrm{{\textstyle isotropic\ curvature}}}\,,\nonumber 
\end{align}
where the parameters and the elastic stress are analogous to the standard
Mindlin-Eringen micromorphic model. The model is well-posed in the
static and dynamical case including when $\mc=0$, see \cite{neff2015relaxed,ghiba2014relaxed}.

The complexity of the classical Mindlin-Eringen micromorphic model
has been decisively reduced featuring basically only symmetric gradient
micro-like variables and the $\Curl$ of the micro-distortion $\p$.
However, the relaxed model is still general enough to include the
full micro-stretch as well as the full Cosserat micro-polar model,
see \cite{neff2014unifying}. Furthermore, well-posedness results
for the static and dynamical cases have been provided in \cite{neff2014unifying}
making decisive use of recently established new coercive inequalities,
generalizing Korn's inequality to incompatible tensor fields \cite{neff2015poincare,neff2002korn,neff2012maxwell,neff2011canonical,bauer2014new,bauer2016dev}.

The relaxed micromorphic model counts 6 constitutive parameters in
the isotropic case ($\me$, $\lle$, $\mh$, $\lh$, $\mc$, $L_{c}$).
The characteristic length $L_{c}$ is intrinsically related to non-local
effects due to the fact that it weights a suitable combination of
first order space derivatives in the strain energy density \eqref{eq:Ener-General}.
For a general presentation of the features of the relaxed micromorphic
model in the anisotropic setting, we refer to \cite{barbagallo2016transparent}.

The associated equations of motion in strong form, obtained by a classical
least action principle take the form (see \cite{madeo2016reflection,madeo2015wave,neff2015relaxed,madeo2014band})
\begin{align}
\rho\,u_{,tt}+\Div[\,\mathcal{I}\,]& =\Div\left[\,\widetilde{\sigma}\,\right], & \eta\,\p_{,tt} & =\widetilde{\sigma}-s-\Curl\,m,\label{eq:Dyn}
\end{align}
where 
\begin{align}
\mathcal{I} & =\overline{\eta}_{1}\,\dev\sym\nablau_{,tt}+\overline{\eta}_{2}\,\skew\nablau_{,tt}+\frac{1}{3}\overline{\eta}_{3}\tr\left(\nablau_{,tt}\right),\nonumber \\
\widetilde{\sigma} & =2\,\me\,\sym\left(\nablau-\p\right)+\lle\,\tr\left(\nablau-\p\right)\id+2\,\mc\,\skew\left(\nablau-\p\right),\\
s & =2\,\mh\,\sym\p+\lh\,\tr\left(\!\p\right)\id,\nonumber \\
m & =\mLc\,\Curl\p.\nonumber 
\end{align}
The fact of adding a gradient micro-inertia in the kinetic energy \eqref{eq:Kinetic} modifies the strong-form PDEs of the relaxed micromorphic model with the addition of the new term $\mathcal{I}$. Of course, boundary conditions would also be modified with respect to the ones presented in \cite{madeo2016reflection,madeo2016first}. The study of the new boundary conditions induced by gradient micro-inertia will be the object of a subsequent paper where the effect of such extra terms on the conservation of energy will also be analyzed.

\begin{figure}[H]
\begin{centering}
\includegraphics[height=5.0cm]{Pictures/RelClassU} \includegraphics[height=5.0cm]{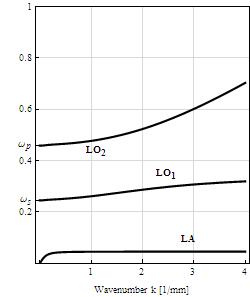}
\includegraphics[height=5.0cm]{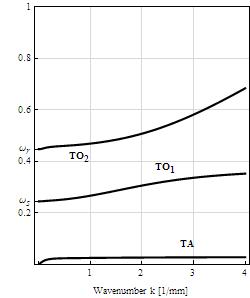} 
\par\end{centering}
\caption{\label{RelClassIn}Dispersion relations $\omega=\omega(k)$ of the
\textbf{relaxed micromorphic model} for the uncoupled (left), longitudinal
(center) and transverse (right) waves with non-vanishing gradient
micro-inertia. Two band-gaps are possible. }
\end{figure}

As it has been shown in previous contributions \cite{madeo2014band,madeo2015wave,madeo2016complete}, the relaxed micromorphic model is able to capture band-gap behaviors thanks to the fact that the acoustic branches have a horizontal asymptote. We show in Figure \ref{RelClassIn} the dispersion relations obtained in previous work which are recovered here setting the gradient micro inertia to be vanishing ($\overline{\eta}=0$). 

\begin{figure}[H]
\begin{centering}
\begin{tabular}{ccccc}
\includegraphics[height=5.0cm]{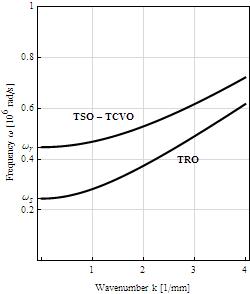} \includegraphics[height=5.0cm]{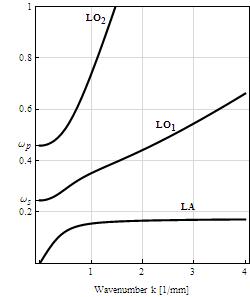}
\includegraphics[height=5.0cm]{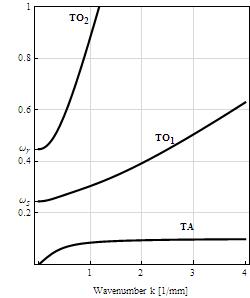}  &  &  &  & \tabularnewline
\end{tabular}
\par\end{centering}
\caption{\label{RelClass}Dispersion relations $\omega=\omega(k)$ of the \textbf{relaxed
micromorphic model} for the uncoupled (left), longitudinal (center)
and transverse (right) waves with vanishing gradient micro-inertia
$\bar{\eta}\protect\neq0$. One complete band-gap is possible. }
\end{figure}

Things are different when adding a gradient micro-inertia $\overline{\eta}\neq0$.
Surprisingly, the combined effect of the free micro-inertia $\eta$
with the gradient micro-inertia can lead to the onset of a second
longitudinal and transverse band gap, due to the fact that the first
longitudinal and transverse acoustic branches ($LO_{1}$ and $TO_{1}$)
are flattened. Moreover, it is possible to notice that the addition
of gradient micro-inertiae $\overline{\eta}_{1}$, $\overline{\eta}_{2}$
and $\overline{\eta}_{3}$ has no effect on the cut-off frequencies,
which only depend on the free micro-inertia $\eta$ (and of course
on the constitutive parameters). 

\begin{figure}[H]
\begin{centering}
\begin{tabular}{ccccc}
 \includegraphics[height=6.0cm]{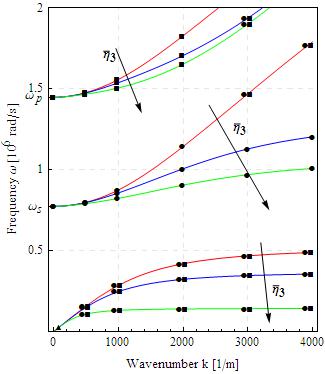}\hspace{1.5cm} \includegraphics[height=6.0cm]{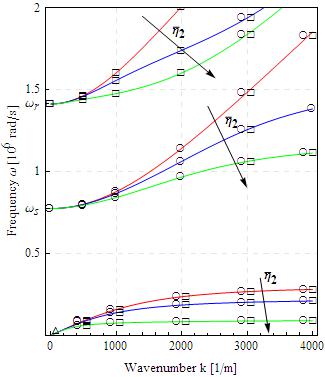}  \tabularnewline
\end{tabular}
\par\end{centering}
\caption{\label{RelClassEig}Dispersion relations $\omega=\omega(k)$ of the
\textbf{relaxed micromorphic model} for longitudinal (left) and transverse
(right) waves with increasing gradient micro-inertia. The markers
indicate the main mode of vibration considering: black triangle $u_{1}$,
black circle $P^{S}$, black square $P^{D}$, empty triangle $u_{\xi}$,
empty circle $P_{(1\xi)}$ and empty square $P_{[1\xi]}$ with $\xi=2,3$.
When two markers are present it means that there is no clear main
mode.}
\end{figure}

In Figure \ref{RelClassEig} we show more explicitly the flattening effect of the gradient inertia parameters on longitudinal and transverse waves. In the same Figure we indicate the main mode of vibration associated to each branch of the dispersion curves. In contrast to Cauchy models, the modes of vibration change when changing the wavenumbers.

In particular, it is possible to notice that the main mode of the acoustic branches
is the longitudinal or transverse displacement (as it is the case
for Cauchy media) only for very small wavenumbers $k$ (large wavelengths).
Increasing the wavenumber (decreasing the wavelength), the longitudinal
and transverse vibrations are characterized by a coupling of the modes
$P^{S}$ and $P^{D}$, and $P_{(1\xi)}$ and $P_{[1\xi]}$ , respectively.
Moreover, it can be seen that the optic branches are characterized by one main
microstructure-related vibrational mode until relatively high values of the
wavenumber $k$. The coupling occurs for higher values of $k$ and
it is strongly influenced by the effects of the gradient micro-inertia.

\section{The micromorphic model with curvature $\rVert\Div\p\lVert^{2}+\rVert\Curl\p\lVert^{2}$\label{DIVCURL}}

We consider now an extension of the relaxed micromorphic model obtained
considering the energy (see \cite{madeo2016complete}): 
\begin{align}
W= & \underbrace{\me\,\lVert\sym\left(\nablau-\p\right)\rVert^{2}+\frac{\lle}{2}\left(\mathrm{tr}\left(\nablau-\p\right)\right)^{2}}_{\mathrm{{\textstyle isotropic\ elastic-energy}}}+\hspace{-0.2cm}\underbrace{\mc\,\lVert\skew\left(\nablau-\p\right)\rVert^{2}}_{\mathrm{{\textstyle rotational\ elastic\ coupling}}}\hspace{-0.1cm}\label{eq:Ener-DivCurl}\\
 & \quad+\underbrace{\mh\,\lVert\sym\p\rVert^{2}+\frac{\lh}{2}\,\left(\mathrm{tr}\p\right)^{2}}_{\mathrm{{\textstyle micro-self-energy}}}+\underbrace{\frac{\mLc}{2}\,\left(\lVert\Div\p\rVert^{2}+\lVert\Curl\p\rVert^{2}\right)}_{\mathrm{{\textstyle augmented\ isotropic\ curvature}}}\,.\nonumber 
\end{align}

The dynamical equilibrium equations are: 
\begin{align}
\rho\,u_{,tt}+\Div[\,\mathcal{I}\,]& =\Div\left[\,\widetilde{\sigma}\,\right], & \eta\,\p_{,tt} & =\widetilde{\sigma}-s-M,\label{eq:DynDivCurl}
\end{align}
where 
\begin{align}
\mathcal{I} & =\overline{\eta}_{1}\,\dev\sym\nablau_{,tt}+\overline{\eta}_{2}\,\skew\nablau_{,tt}+\frac{1}{3}\overline{\eta}_{3}\tr\left(\nablau_{,tt}\right),\nonumber \\
\widetilde{\sigma} & =2\,\me\,\sym\left(\nablau-\p\right)+\lle\,\tr\left(\nablau-\p\right)\id+2\,\mc\,\skew\left(\nablau-\p\right),\\
s & =2\,\mh\,\sym\p+\lh\,\tr\left(\!\p\right)\id,\nonumber \\
M & =-\mLc\,\underbrace{\left(\nabla\left(\Div\p\right)-\Curl\Curl\p\right)}_{=\Div\nabla\p=\Delta\p}.\nonumber 
\end{align}
Note that the structure of the equation is equivalent to the one obtained
in the standard micromorphic model with curvature $\frac{1}{2}\rVert\nablap\lVert^{2}$,
see equation \eqref{eq:DynMin} in section \ref{sec:Min}.

\begin{figure}[H]
\begin{centering}
\begin{tabular}{ccccc}
\includegraphics[height=5.0cm]{Pictures/RelClassU} \includegraphics[height=5.0cm]{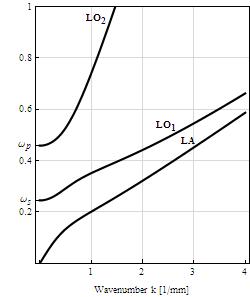}
\includegraphics[height=5.0cm]{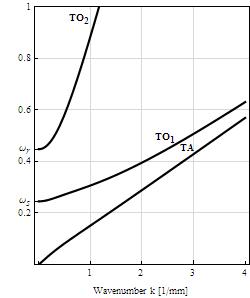}  &  &  &  & \tabularnewline
\end{tabular}
\par\end{centering}
\caption{\label{RelCurlDiv}Dispersion relations $\omega=\omega(k)$ of the
\textbf{relaxed micromorphic model with curvature $\rVert\Div\p\lVert^{2}+\rVert\Curl\p\lVert^{2}$}
for the uncoupled (left), longitudinal (center) and transverse (right)
waves with vanishing gradient micro-inertia. No band-gap is possible.}
\end{figure}

\begin{figure}[H]
\begin{centering}
\begin{tabular}{ccccc}
\includegraphics[height=5.0cm]{Pictures/RelClassU} \includegraphics[height=5.0cm]{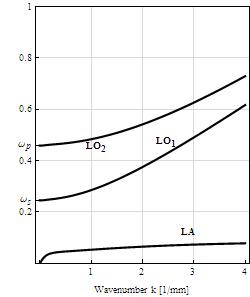}
\includegraphics[height=5.0cm]{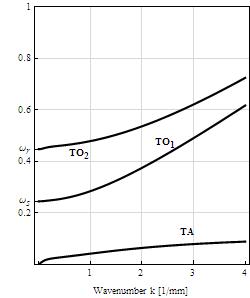}  &  &  &  & \tabularnewline
\end{tabular}
\par\end{centering}
\caption{\label{RelCurlDivIn}Dispersion relations $\omega=\omega(k)$ of the
\textbf{relaxed micromorphic model with curvature $\rVert\Div\p\lVert^{2}+\rVert\Curl\p\lVert^{2}$}
for the uncoupled (left), longitudinal (center) and transverse (right)
waves with non-vanishing gradient micro-inertia. One band-gap is possible. }
\end{figure}

We present the dispersion relations obtained with a vanishing
gradient inertia (Figure \ref{RelCurlDiv}) and for a non-vanishing
gradient micro-inertia (Figure \ref{RelCurlDivIn}). We conclude that when
considering the model with micromorphic medium with $\rVert\Div\p\lVert^{2}$
+ $\rVert\Curl\p\lVert^{2}$ with vanishing gradient micro-inertia,
there always exist waves which propagate inside the considered medium
independently of the value of frequency even if considering separately
longitudinal, transverse and uncoupled waves. On the other hand, switching
on the gradient inertia it is possible to obtain a total band-gap.

\section{The micromorphic model with curvature $\rVert\Div\p\lVert^{2}$}

The isotropic micromorphic model with $\rVert\Div\p\lVert^{2}$ is
yet another variant of the classical relaxed micromorphic model (see
\cite{madeo2016complete}) with energy: 
\begin{align}
W= & \underbrace{\me\,\lVert\sym\left(\nablau-\p\right)\rVert^{2}+\frac{\lle}{2}\left(\mathrm{tr}\left(\nablau-\p\right)\right)^{2}}_{\mathrm{{\textstyle isotropic\ elastic-energy}}}+\hspace{-0.1cm}\underbrace{\mc\,\lVert\skew\left(\nablau-\p\right)\rVert^{2}}_{\mathrm{{\textstyle rotational\ elastic\ coupling}}}\hspace{-0.1cm}\label{eq:Ener-Div}\\
 & \quad+\underbrace{\mh\,\lVert\sym\p\rVert^{2}+\frac{\lh}{2}\,\left(\mathrm{tr}\p\right)^{2}}_{\mathrm{{\textstyle micro-self-energy}}}+\hspace{-0.2cm}\underbrace{\frac{\mLd}{2}\,\lVert\Div\p\rVert^{2}}_{\mathrm{{\textstyle isotropic\ curvature}}}\,.\nonumber 
\end{align}
The dynamical equilibrium equations are: 
\begin{align}
\rho\,u_{,tt}+\Div[\,\mathcal{I}\,]& =\Div\left[\,\widetilde{\sigma}\,\right], & \eta\,\p_{,tt} & =\widetilde{\sigma}-s-M,\label{eq:DynDiv}
\end{align}
where 
\begin{align}
\mathcal{I} & =\overline{\eta}_{1}\,\dev\sym\nablau_{,tt}+\overline{\eta}_{2}\,\skew\nablau_{,tt}+\frac{1}{3}\overline{\eta}_{3}\tr\left(\nablau_{,tt}\right),\nonumber \\
\widetilde{\sigma} & =2\,\me\,\sym\left(\nablau-\p\right)+\lle\,\tr\left(\nablau-\p\right)\id+2\,\mc\,\skew\left(\nablau-\p\right),\\
s & =2\,\mh\,\sym\p+\lh\,\tr\left(\!\p\right)\id,\nonumber \\
M & =-\mLc\,\nabla\left(\Div\p\right).\nonumber 
\end{align}

\begin{figure}[H]
\begin{centering}
\begin{tabular}{ccccc}
\includegraphics[height=5.0cm]{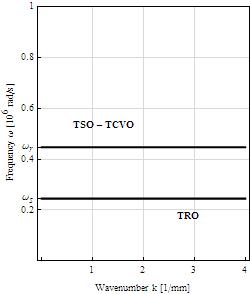} \includegraphics[height=5.0cm]{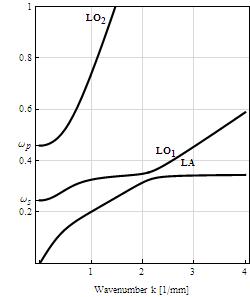}
\includegraphics[height=5.0cm]{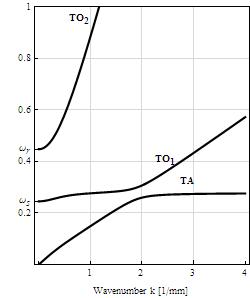}  &  &  &  & \tabularnewline
\end{tabular}
\par\end{centering}
\caption{\label{RelDiv}Dispersion relations $\omega=\omega(k)$ of the \textbf{relaxed
micromorphic model with curvature $\rVert\Div\p\lVert^{2}$} for the
uncoupled (left), longitudinal (center) and transverse (right) waves
with vanishing gradient micro-inertia. No band-gap is possible.}
\end{figure}

\begin{figure}[H]
\begin{centering}
\begin{tabular}{ccccc}
\includegraphics[height=5.0cm]{Pictures/IntU} \includegraphics[height=5.0cm]{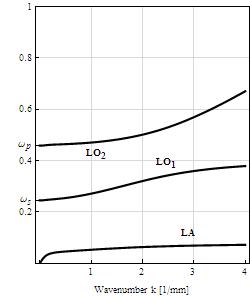}
\includegraphics[height=5.0cm]{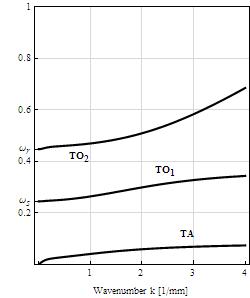}  &  &  &  & \tabularnewline
\end{tabular}
\par\end{centering}
\caption{\label{RelDivIn}Dispersion relations $\omega=\omega(k)$ of the \textbf{relaxed
micromorphic model with curvature $\rVert\Div\p\lVert^{2}$} for the
uncoupled (left), longitudinal (center) and transverse (right) waves
with non-vanishing gradient micro-inertia. Two band-gaps are possible.}
\end{figure}

We present the \textbf{dispersion relations} obtained with a non-vanishing
gradient inertia (Figure \ref{RelDivIn}) and for a vanishing gradient
inertia (Figure \ref{RelDiv}). In the Figures we consider uncoupled
waves (a), longitudinal waves (b) and transverse waves (c). Even in
this case, when considering the micromorphic model with only $\rVert\Div\p\lVert^{2}$
with vanishing gradient inertia, there always exist waves which propagate
inside the considered medium independently of the value of the frequency
and the uncoupled waves assume a peculiar behavior in which the frequency
is independent of the wavenumber k. On the other hand, when switching
on the gradient inertia, a behavior analogous to the relaxed micromorphic
model appears: it is possible to model the onset of two complete band-gaps.
The uncoupled waves remain unaffected by the introduction of the gradient
micro-inertia and they keep their characteristic of being independent
of the wavenumber in strong contrast to what happen for the relaxed
micromorphic model in which the uncoupled waves are dispersive. 

\section{The standard Mindlin-Eringen model \label{sec:Min}}

In this section. we discuss the effect on the Mindlin-Eringen of the
addition of the gradient micro-inertia $\overline{\eta}\lVert\nablau_{,t}\rVert^{2}$
to the classical terms $\rho\lVert u_{,t}\rVert^{2}+\eta\lVert\p_{,t}\rVert^{2}$.
We recall that the strain energy density for this model in the isotropic
case takes the form: 
\begin{align}
W= & \underbrace{\me\,\lVert\sym\left(\nablau-\p\right)\rVert^{2}+\frac{\lle}{2}\left(\mathrm{tr}\left(\nablau-\p\right)\right)^{2}}_{\mathrm{{\textstyle isotropic\ elastic-energy}}}+\hspace{-0.1cm}\underbrace{\mc\,\lVert\skew\left(\nablau-\p\right)\rVert^{2}}_{\mathrm{{\textstyle rotational\ elastic\ coupling}}}\hspace{-0.1cm}\label{eq:Ener-Mindlin}\\
 & \quad+\underbrace{\mh\,\lVert\sym\p\rVert^{2}+\frac{\lh}{2}\,\left(\mathrm{tr}\p\right)^{2}}_{\mathrm{{\textstyle micro-self-energy}}}+\hspace{-0.2cm}\underbrace{\frac{\mLc}{2}\,\lVert\nabla\,\p\rVert^{2}}_{\mathrm{{\textstyle isotropic\ curvature}}}\,,\nonumber 
\end{align}
The dynamical equilibrium equations are: 
\begin{align}
\rho\,u_{,tt}+\Div[\,\mathcal{I}\,]& =\Div\left[\,\widetilde{\sigma}\,\right], & \eta\,\p_{,tt} & =\widetilde{\sigma}-s-M,\label{eq:DynMin}
\end{align}
where 
\begin{align}
\mathcal{I} & =\overline{\eta}_{1}\,\dev\sym\nablau_{,tt}+\overline{\eta}_{2}\,\skew\nablau_{,tt}+\frac{1}{3}\overline{\eta}_{3}\tr\left(\nablau_{,tt}\right),\nonumber \\
\widetilde{\sigma} & =2\,\me\,\sym\left(\nablau-\p\right)+\lle\,\tr\left(\nablau-\p\right)\id+2\,\mc\,\skew\left(\nablau-\p\right),\\
s & =2\,\mh\,\sym\p+\lh\,\tr\left(\!\p\right)\id,\nonumber \\
M & =-\mLc\,\underbrace{\Div\nabla\p}_{=\Delta\p}.\nonumber 
\end{align}

\begin{figure}[H]
\begin{centering}
\includegraphics[height=5.0cm]{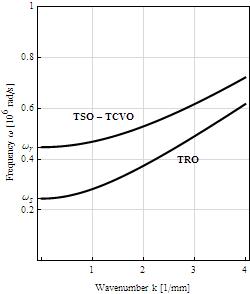} \includegraphics[height=5.0cm]{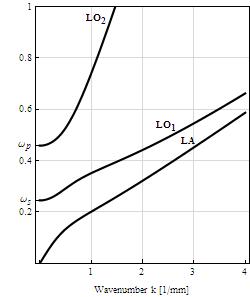}
\includegraphics[height=5.0cm]{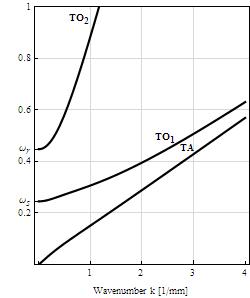} 
\par\end{centering}
\caption{\label{Min}Dispersion relations $\omega=\omega(k)$ of the \textbf{standard
Mindlin-Eringen micromorphic model} for the uncoupled (left), longitudinal
(center) and transverse (right) waves with vanishing gradient micro-inertia. No band-gap is possible.}
\end{figure}

\begin{figure}[H]
\begin{centering}
\begin{tabular}{ccccc}
\includegraphics[height=5.0cm]{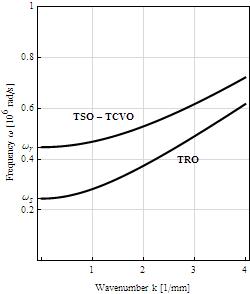} \includegraphics[height=5.0cm]{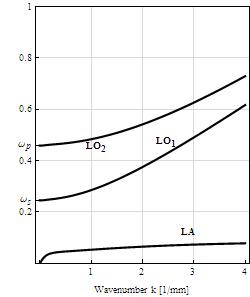}
\includegraphics[height=5.0cm]{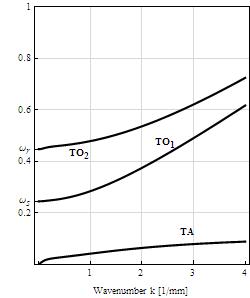}  &  &  &  & \tabularnewline
\end{tabular}
\par\end{centering}
\caption{\label{MinIn}Dispersion relations $\omega=\omega(k)$ of the \textbf{standard
Mindlin-Eringen micromorphic model} for the uncoupled (left), longitudinal
(center) and transverse (right) waves with non-vanishing gradient
micro-inertia. One band-gap is possible.}
\end{figure}

Recalling the results of \cite{madeo2014band}, we remark that when
the gradient micro-inertia is vanishing ($\overline{\eta}_{1}=\overline{\eta}_{2}=\overline{\eta}_{3}=0$)
the Mindlin-Eringen model does not allow the description of band-gaps
(see Fig.$\,$\ref{Min}), due to the presence of straight acoustic
waves. On the other hand, when switching on the parameters $\overline{\eta}_{2}$
and $\overline{\eta}_{3}$ , some optic branches are flattened, so
that the first band-gap can be created (see Fig.$\,$\ref{MinIn}). The analogous
case for the relaxed micromorphic model (Fig.$\,$\ref{RelClass})
allowed instead for the description of 2 band gaps.

As already pointed out and as shown in \cite{madeo2016complete},
the classical Mindlin-Eringen model can be considered to be equivalent
to the relaxed micromorphic model with curvature $\rVert\Div\p\lVert^{2}+\rVert\Curl\p\lVert^{2}$.

\section{The internal variable model}

\begin{figure}[H]
\begin{centering}
\includegraphics[height=5.0cm]{Pictures/IntU} \includegraphics[height=5.0cm]{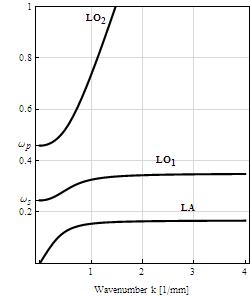}
\includegraphics[height=5.0cm]{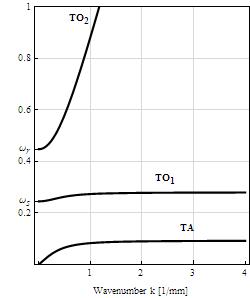} 
\par\end{centering}
\caption{\label{Int}Dispersion relations $\omega=\omega(k)$ of the \textbf{internal
variable model} for the uncoupled (left), longitudinal (center) and
transverse (right) waves with vanishing gradient micro-inertia. Two band-gaps are possible but non-local effects cannot be described.}
\end{figure}

Figure \ref{Int} shows the behavior of the addition of the gradient
micro-inertia $\overline{\eta}\lVert\nablau_{,t}\rVert^{2}$ in the
internal variable model. We recall (see \cite{neff2014unifying})
that the energy for the internal variable model does not include higher
space derivatives of the micro-distortion tensor $\p$ and, in the
isotropic case, takes the form: 
\begin{align}
W= & \underbrace{\me\,\lVert\sym\left(\nablau-\p\right)\rVert^{2}+\frac{\lle}{2}\left(\mathrm{tr}\left(\nablau-\p\right)\right)^{2}}_{\mathrm{{\textstyle isotropic\ elastic-energy}}}+\hspace{-0.1cm}\underbrace{\mc\,\lVert\skew\left(\nablau-\p\right)\rVert^{2}}_{\mathrm{{\textstyle rotational\ elastic\ coupling}}}\hspace{-0.1cm}\label{eq:Ener-Int}\\
 & \quad+\underbrace{\mh\,\lVert\sym\p\rVert^{2}+\frac{\lh}{2}\,\left(\mathrm{tr}\p\right)^{2}}_{\mathrm{{\textstyle micro-self-energy}}}\,,\nonumber 
\end{align}
The dynamical equilibrium equations are: 
\begin{align}
\rho\,u_{,tt}+\Div[\,\mathcal{I}\,]& =\Div\left[\,\widetilde{\sigma}\,\right], & \eta\,\p_{,tt} & =\widetilde{\sigma}-s,\label{eq:DynInt}
\end{align}
where 
\begin{align}
\mathcal{I} & =\overline{\eta}_{1}\,\dev\sym\nablau_{,tt}+\overline{\eta}_{2}\,\skew\nablau_{,tt}+\frac{1}{3}\overline{\eta}_{3}\tr\left(\nablau_{,tt}\right),\nonumber \\
\widetilde{\sigma} & =2\,\me\,\sym\left(\nablau-\p\right)+\lle\,\tr\left(\nablau-\p\right)\id+2\,\mc\,\skew\left(\nablau-\p\right),\\
s & =2\,\mh\,\sym\p+\lh\,\tr\left(\!\p\right)\id.\nonumber 
\end{align}

\begin{figure}[H]
\begin{centering}
\includegraphics[height=5.0cm]{Pictures/IntU} \includegraphics[height=5.0cm]{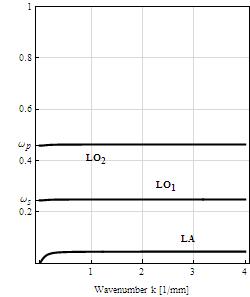}
\includegraphics[height=5.0cm]{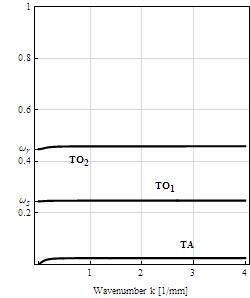} 
\par\end{centering}
\caption{\label{IntIn}Dispersion relations $\omega=\omega(k)$ of the \textbf{internal
variable model} for the uncoupled (left), longitudinal (center) and
transverse (right) waves with non-vanishing gradient micro-inertia. Three band-gap are possible but non-local effects cannot be described. The overall trend of the dispersion curves is unrealistic for the great majority of metamaterials.}
\end{figure}

We present the dispersion relations obtained for the internal variable
model with a non-vanishing gradient inertia (Fig.$\,$\ref{IntIn})
and for a vanishing gradient inertia (Fig.$\,$\ref{Int}). We start
noticing in Fig.$\,$\ref{Int} that the internal variable model with
vanishing gradient micro-inertia allows for the description of two
complete band-gap even if it is not able to account for the presence
of non-localities in metamaterials. Moreover, by direct observation
of Fig.$\,$\ref{IntIn}, we can notice that, when switching on the
gradient micro-inertia, suitably choosing the relative position of
$\omega_{r}$ and $\omega_{p}$, the internal variable model allows
to account for 3 band gaps. We thus have an extra band-gap with respect
to the case with vanishing gradient inertia (Fig.$\,$\ref{Int})
and to the analogous case for the relaxed micromorphic model (see
Fig.$\,$\ref{RelClassIn}), but we are not able to consider non-local
effects. The fact of excluding the possibility of describing non-local
effects in metamaterials can be sometimes too restrictive. For example,
flattening the curve which originates from $\omega_{r}$ and which
is associated to rotational modes of the microstructure is nonphysical
for the great majority of metamaterials.

\section{Conclusions}

In this paper we make a review of some of the available isotropic,
linear-elastic, enriched continuum models for the description of the
dynamical behavior of metamaterials. We show that the relaxed micromorphic
model previously introduced by the authors is the only non-local enriched
model which is able to describe band-gaps when considering a kinetic
energy independently accounting for micro and micro motions. As far
as an inertia term which couples the micro-motions to the macroscopic
motions of the unit cell is introduced, also other non-local models
exhibit the possibility of describing band-gap behaviors. Nevertheless,
the relaxed micromorphic model is still the more effective one to
describe (multiple) band-gaps and non-local effects in a realistic
way. In fact, even with the addition of the new micro-inertia term,
the relaxed model is able to account for the description of two band-gaps,
in contrast with to the single band-gap allowed by the Mindlin-Eringen
model. The micromorphic model with curvature $\rVert\Div\p\lVert^{2}$
also allows for the description of two band-gaps when considering
an augmented kinetic energy, but the uncoupled waves are forced to
be non-dispersive: this fact can be considered to be a limitation
for the realistic description of a wide class of band-gap metamaterials.
Finally, the internal variable model with the new kinetic energy terms
allow for the description of up to three band gaps. Nevertheless,
the overall trends shown by the dispersion curves turn to be quite
unrealistic due to the fact that all the branches of the dispersion
curves show very low or no dispersion at all.

\section{Acknowledgments}

The work of Ionel-Dumitrel Ghiba was supported by a grant of the Romanian National Authority for Scientific Research and Innovation, CNCS-UEFISCDI, project number PN-II-RU-TE-2014-4-0320. Angela Madeo thanks INSA-Lyon for the funding of the BQR 2016 \textquotedbl{}Caractérisation
mécanique inverse des métamatériaux: modélisation, identification
expérimentale des paramètres et évolutions possibles\textquotedbl{},
as well as the CNRS-INSIS for the funding of the PEPS project.

\bibliographystyle{plain}
\bibliography{library}

\end{document}